%% file: Main.tex
\documentclass[a4paper,11pt]{article}
\pdfoutput=1 % if your are submitting a pdflatex (i.e. if you have
\usepackage{jinstpub} % for details on the use of the package, please
\usepackage{verbatim}
\usepackage{siunitx}

\usepackage{amsmath}
\newcommand*\subtxt[1]{_{\textnormal{#1}}}
\DeclareRobustCommand\_{\ifmmode\expandafter\subtxt\else\textunderscore\fi}

\usepackage{caption} 
\usepackage{lineno}
\title{\boldmath Picosecond Avalanche Detector - working principle and gain measurement with a proof-of-concept prototype}
\author[a,b,1]{L. Paolozzi,\note{Corresponding author.}} 
\author[a]{M. Munker,}
\author[a]{R. Cardella,}
\author[a]{M. Milanesio,}
\author[a]{Y. Gurimskaya,}
\author[a]{F. Martinelli,}
\author[a,b]{A. Picardi,}
\author[c]{H. Rücker,}
\author[c]{A. Trusch,}
\author[a]{P. Valerio,}

\author[a]{F. Cadoux,}
\author[a]{R. Cardarelli,}
\author[a]{S. Débieux,}
\author[a]{Y. Favre,}
\author[a]{C. A. Fenoglio,}
\author[a]{D. Ferrere,}
\author[a]{S. Gonzalez-Sevilla,}
\author[a,b]{R. Kotitsa,}
\author[a]{C. Magliocca,}
\author[a]{T. Moretti,}
\author[a,b]{M. Nessi,}
\author[a]{A. Pizarro Medina,}
\author[a]{J. Sabater Iglesias,}
\author[a]{J. Saidi,}
\author[a]{M. Vicente Barreto Pinto,}
\author[b]{S. Zambito}
\author[a]{and G. Iacobucci}
\affiliation[a]{D\'epartement de Physique Nucl\'eaire et Corpusculaire (DPNC),
University of Geneva, 24 Quai Ernest-Ansermet, CH-1205 Geneva 4, Switzerland}

\affiliation[b]{CERN, CH-1211 Geneva 23, Switzerland}

\affiliation[c]{IHP — Leibniz-Institut für innovative Mikroelektronik, Im Technologiepark 25, Frankfurt (Oder), Germany}

\emailAdd{lorenzo.paolozzi@cern.ch}

\abstract{The Picosecond Avalanche Detector is a multi-junction silicon pixel detector based on a $ \mathrm{(NP)_{drift}(NP)_{gain}} $ structure, devised to enable charged-particle tracking with high spatial resolution and picosecond time-stamp capability. It uses a continuous junction deep inside the sensor volume to amplify the primary charge produced by ionizing radiation in a thin absorption layer. The signal is then induced by the secondary charges moving inside a thicker drift region. A proof-of-concept monolithic  prototype, consisting of a matrix of hexagonal pixels with 100 µm  pitch, has been produced using the 130 nm SiGe BiCMOS process by IHP microelectronics. Measurements on probe station and with a $ ^{55} $Fe X-ray source show that the prototype is functional and  displays avalanche gain up to a maximum electron gain of 23.  A study of the avalanche characteristics, corroborated by TCAD simulations, indicates that  space-charge effects due to the large primary charge produced by the conversion of X-rays from the $ ^{55} $Fe source limits the effective gain.}


\begin{document}
\maketitle 
\flushbottom

\input{1_Introduction}

\input{2_Prototypes}

\input{3_TCAD_Simulations}

\input{4_Probe_Station}

\input{5_Climate_Chamber}

\input{6_Conclusions}

\acknowledgments

The prototyping of the PicoAD was funded by the  EU H2020 ATTRACT MONPICOAD project under grant agreement 222777, with support by INNOGAP funds from the University of Geneva. The test and characterization of the prototypes was done in the context of the H2020 ERC Advanced Grant MONOLITH, grant ID: 884447. The authors wish to thank Coralie Husi, Javier Mesa, Gabriel Pelleriti and all the technical staff of the University of Geneva and IHP microelectronics. The authors acknowledge the support of EUROPRACTICE in providing design tools and MPW fabrication services.
%\paragraph{Note added.} This is also a good position for notes added
%after the paper has been written.


\newpage
\bibliographystyle{unsrt}
\bibliography{bibliography.bib}

\end{document}

%% file: 1_Introduction.tex
\section{Introduction}
\label{sec:intro}

The monolithic detector presented in this paper, the  Picosecond Avalanche Detector (PicoAD) \cite{PicoADpatent}, is a multi-junction silicon pixel sensor  for the detection of light and ionizing radiation, developed in the framework of the  H2020 ERC Advanced MONOLITH project \cite{monolith}. It is %operated in proportional avalanche mode and 
devised to combine the space resolution and high detection efficiency of PIN sensors with picosecond time resolution by enhancing the timing performance by avalanche multiplication in a continuous gain layer implanted deep in the depleted region.

\subsection{Design challenges of avalanche diodes for picosecond timing measurement\\ with ionizing radiation}

Avalanche diodes have been recently used to detect ionizing radiation, exploiting the signal amplification provided by the impact ionization to increase the signal-to-noise ratio and improve the detector time resolution \cite{HAUGER1994362}\cite{CARULLA2019373}\cite{HGTD1}. This approach was successful in achieving tens of picoseconds time resolution with Minimum Ionizing Particles (MIPs), but some aspects of the sensor design present challenges: the reach-through structure typically used in N-on-P diodes to localize the gain layer underneath the pixel is susceptible to the strong variations of the electric field at the pixel edge, requiring to introduce discontinuities of the gain layer in the inter-pixel regions, which could impact the sensor detection efficiency and time resolution. This problem is often mitigated by increasing the pixel size to reduce the fraction of the sensor inactive area \cite{HGTD1}\cite{SADROZINSKI201618}. Another challenge to the development of this technology is the presence of charge-collection noise \cite{TesiPaolozzi}, which produces an intrinsic time jitter during the collection of the signal charge, generated by the fluctuations of the charge profile along the path of the primary ionization. This effect can be reduced by thinning the avalanche diode, \cite{CARULLA2019373}\cite{TesiPaolozzi}\cite{werner} at the cost of increasing the detector capacitance and consequently the time jitter of the front-end electronics. 
Charge-collection noise thus makes it difficult to achieve time resolutions below 20 ps for MIPs with present avalanche diodes. The reason is mostly due to the fact that the region where the primary charge is produced coincides with the region where the secondary charges drift and contributes to most of the induced signal.

\section{Working principle of the  Picosecond Avalanche Detector}

\begin{figure}[htbp]
\centering % \begin{center}/\end{center} takes some additional vertical space
\includegraphics[width=.9\textwidth,trim=0 0 0 0,clip]{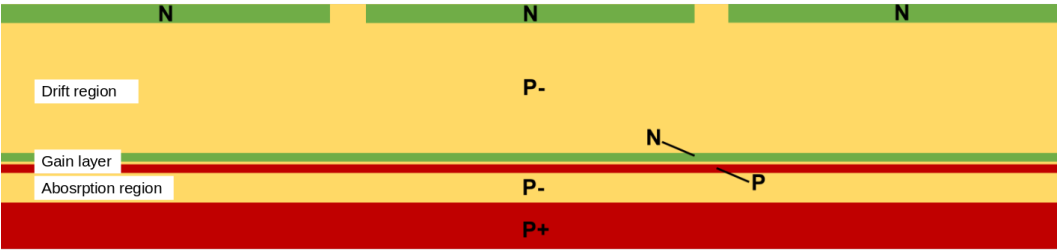}
% "\includegraphics" from the "graphicx" permits to crop (trim+clip)
% and rotate (angle) and image (and much more)
\caption{\label{fig:geometry} Simplified cross section of the PicoAD detector. The sensor presents N-type pixels on a high-resistivity epitaxial layer with boron background. A second junction, deep inside the sensor volume, is used to produce a continuous avalanche gain layer. The epitaxial layer and the deep junction are operated in full depletion.}
\end{figure}

The PicoAD introduces design flexibility to the avalanche-based sensors by moving the gain layer away from the pixel matrix and separating the region where the primary charge is produced and amplified from the region dedicated to the signal induction on the readout electrodes. Figure \ref{fig:geometry} shows the cross section of the PicoAD. The detector is characterized by a fully depleted multi-junction (NP)$_{pixel}$(NP)$_{gain}$ structure, which provides a new degree of freedom for the engineering of the electric field. Indeed, when the bias voltage is applied, the sensor depletion region expands from the two junctions that operate in inverse polarization: the one close to the top surface (pixel junction) serves to isolate the pixels, and the one close to the bottom edge of the depletion region (gain junction) provides an electric field large enough to generate impact ionization and, therefore, avalanche gain. The central, direct junction is isolated by the other two junctions until full depletion is achieved. 

When crossed by ionizing radiation, the primary holes generated inside the region that extends between the pixel and gain junctions will drift towards the gain junction, featuring little or no amplification in silicon, while the primary electrons generated below the gain junction will drift upwards to be amplified, producing secondary charges. This behavior allows identifying three main regions:
\begin{itemize}
    \item The primary \textit{absorption region}, constituted by the bottom P- doped region, near the backside contact. This region does not produce impact ionization itself, but the $e^-$ signal generated here will be amplified by the PicoAD.
    \item The uniform and continuous deep NP junction is the \textit{gain layer}, operated in proportional mode and therefore free from dark noise counts when detecting ionizing radiation.
    \item The \textit{drift region}, which corresponds to the thicker top P- doped region, between the pixels and the gain layer. Primary charges and secondary electrons moving in this region induce most of the signal current in the pixel above, according to the Schockley-Ramo theorem \cite{Shockley}\cite{Ramo}.
\end{itemize}

This sensor structure offers multiple advantages: 
\begin{itemize}
\item[{\it i)}]
First and foremost, the gain layer is a uniform implant far from the pixels, therefore it is less subject to strong variations of the electric field, and can grant uniform signal amplification on all the detector surface. This characteristic enables the possibility to reduce the pixel size with less impact on the avalanche gain performance, as long as the inter-pixel distance is smaller than or comparable to the drift-region thickness. 
\item[{\it ii)}]Another advantage is that the absorption region can be made arbitrarily thin to reduce the charge-collection noise without significantly increasing the pixel capacitance. This is a fundamental feature because it allows exploiting the advances in fast, low-noise electronics to boost the signal-to-noise ratio and achieve picosecond time resolution. 
\item[{\it iii)}]
Finally, the design of the pixel matrix does not depend on the presence of the gain layer. This makes the R\&D simpler because it allows integrating the PicoAD gain layer in an existing sensor design.
\end{itemize}

\begin{figure}[htbp]
\centering % \begin{center}/\end{center} takes some additional vertical space
\includegraphics[width=.99\textwidth,trim=0 0 0 0,clip]{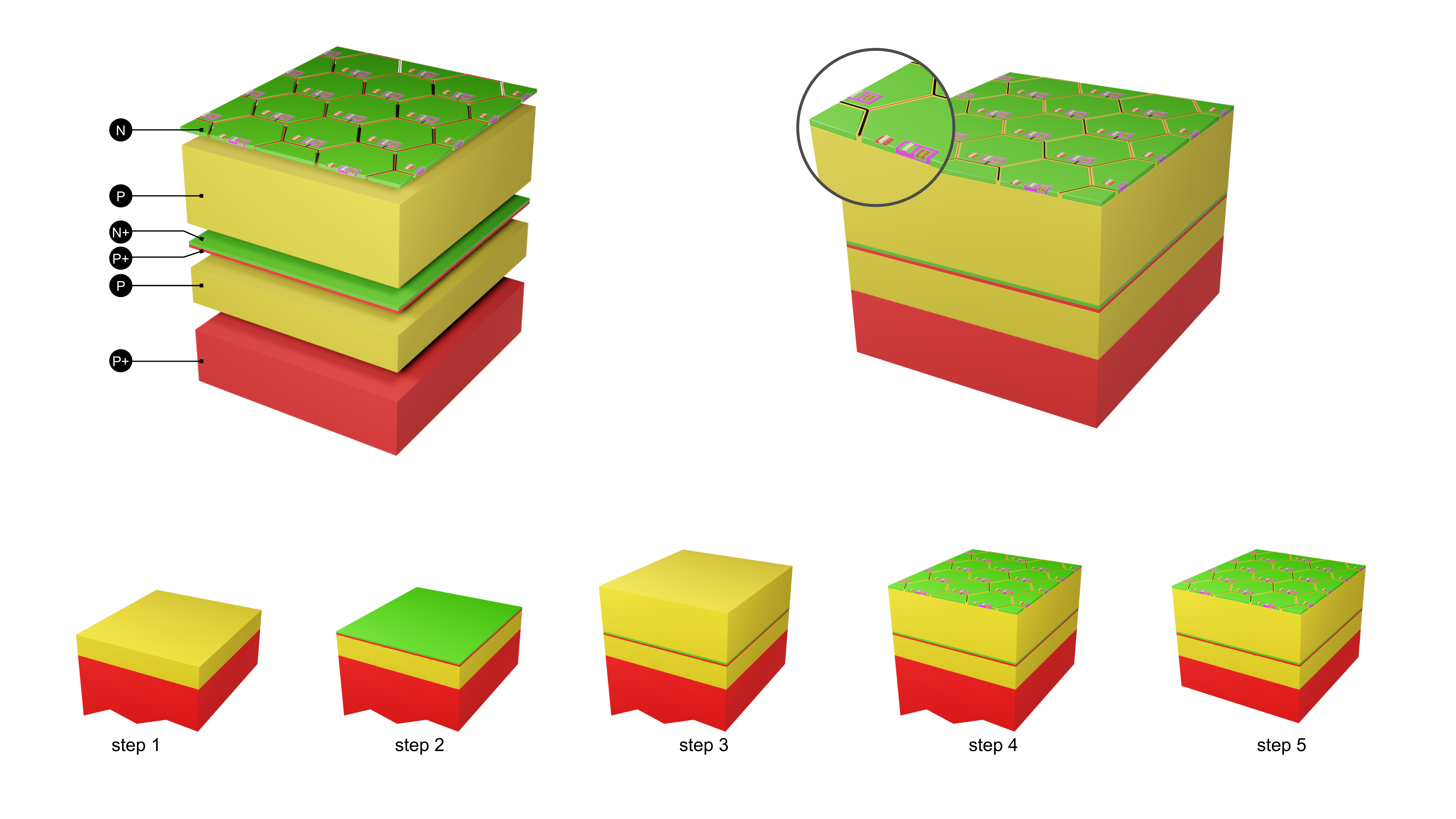}
% "\includegraphics" from the "graphicx" permits to crop (trim+clip)
% and rotate (angle) and image (and much more)
\caption{\label{fig:manufacturing} Manufacturing process of the PicoAD. ~Step 1: a high-resistivity epitaxial (absorption) layer is grown on a low resistivity, boron-doped substrate. Step 2: a uniform NP gain  layer is implanted. Step 3: a second epitaxial layer is produced, which corresponds to the sensor drift region. Step 4: the top surface is processed to produce the pixel matrix; in the case of a monolithic implementation of the PicoAD, this is a full-CMOS processing. Step 5:  the substrate can be thinned from the backside and metallized.}
\end{figure}

The distinctive characteristics of the PicoAD is its multi-junction structure.
It requires the implantation of uniform, deep NP doping profiles to generate the gain layer close to the backside surface of the sensor depleted region.  Given the thin depletion region required to achieve picosecond time resolution, a possible approach to manufacturing the PicoAD is producing special epitaxial wafers that integrate the gain structure before the surface processing of the wafer.
The procedure adopted for the manufacturing of the proof-of-concept prototype is depicted in Figure \ref{fig:manufacturing}

%% file: 2_Prototypes.tex
\section{The proof-of-concept PicoAD prototypes}
\label{sec:chip}

Proof-of-concept PicoAD prototypes were produced in a monolithic implementation by IHP. The manufacturing of the prototypes was divided in wafer preparation and CMOS processing using their SG13G2 130nm SiGe BiCMOS  process. 
\\

{\it Wafer preparation}

The PicoAD  was produced on boron-doped substrates with a resistivity of 0.1 $\Omega$cm. To guarantee the uniformity of the electric field in the gain layer, the drift region of the PicoAD should be thicker than the inter-pixel distance of the matrix. This condition was not compatible with a full depletion of the sensor using the epitaxial process with the minimum boron concentration of $ 3\cdot 10^{14}$ cm$ ^{-3} $ available at IHP at the time of manufacturing. For this reason,
the prototype wafer  produced was made by a 5 µm thick epitaxial layer under the gain layer (absorption region) and a 10 µm thick epitaxial layer above the gain layer (drift region).
The thickness of the drift region of this proof-of-concept PicoAD prototype is thus  sub-optimal, and may lead to a reduction of the avalanche gain and a distortion of the weighting field responsible for signal induction in the inter-pixel region. 

Four implantation doses were used to form the gain layer (dose 1 being the lowest, dose 4 the highest), targeting a sensor working point at a bias voltage between 100 V and 200 V. In all cases the dose of the P-type implant was set at 60\% of the dose of the N-type implant, to guarantee a reduction of the electric field in the absorption region when full depletion is reached.

One of the risks associated with the manufacturing of the multi-junction structure is the presence of a direct junction in the transition between the drift and the gain regions. During sensor operation, when the epitaxial layer is fully depleted, this region is emptied of all free charges. But this is not the case during the ramp-up of the sensor bias: at this stage, if the gain layer implantation reaches the edge of the sensor, a conductive path with the substrate may be formed where the silicon is cut, producing an excess current in the device. To avoid this effect it was decided to use an implantation mask to guarantee that the gain layer ended at 80 µm from the chip edge.

The wafers were produced using the process that was available at IHP, which is a dichlorsilane-based process at 850°C, optimized for sub-micron epitaxial processing. This process could be used to make proof-of-concept prototypes, but it is not suitable for producing large volume of wafers as it is time consuming and has non-negligible risk of introducing defects on the surface of the thicker epitaxial drift layer.
\\

{\it CMOS processing}

\begin{figure}[htbp]
\centering % \begin{center}/\end{center} takes some additional vertical space
\includegraphics[width=.55\textwidth,trim=0 0 0 0,clip]{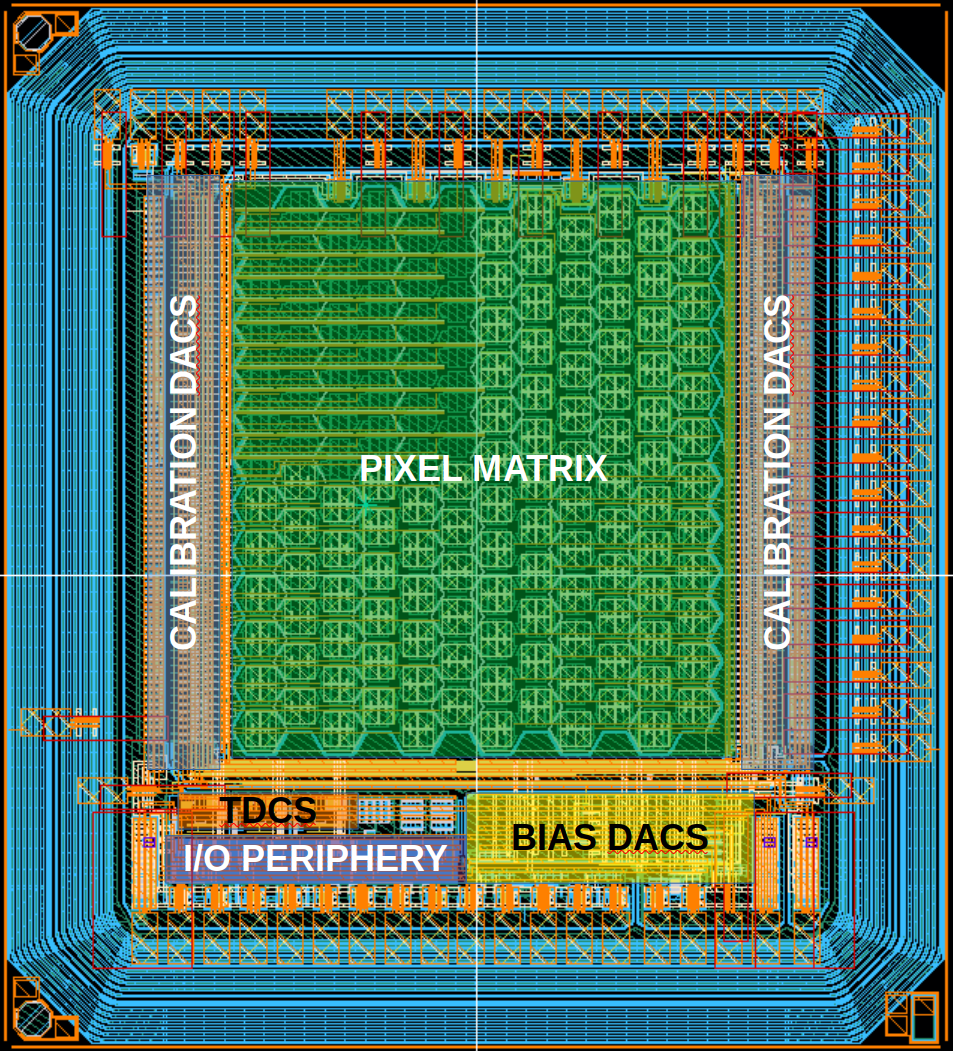}
% "\includegraphics" from the "graphicx" permits to crop (trim+clip)
% and rotate (angle) and image (and much more)
\caption{\label{fig:floorplan} Layout of the monolithic silicon pixel ASIC used to test the PicoAD proof-of-concept prototypes. A detailed description of the ASIC layout and operation is reported in \cite{Iacobucci:2021ukp}.}
\end{figure}

Due to its thin absorption layer, the PicoAD generates ultra-fast signals that are smaller than those produced by  typical avalanche diodes used to detect ionizing particles. For this reason, this sensor should be integrated with ultra-fast, low noise front-end electronics. The best analog performance in terms of speed and noise of the pre-amplification stage can be achieved using Silicon-Germanium Heterojunction Bipolar Transistors (SiGe HBT) \cite{SiGe_tech}. For this reason, the monolithic silicon pixel layout already designed and produced in 130 nm SiGe BiCMOS of IHP for the H2020 ATTRACT MonPicoAD project\footnote{For the ATTRACT MonPicoAD project, this layout was  produced on standard wafers without internal gain and characterised in a testbeam with 180 GeV pions, showing full efficiency and time resolution of 36 ps \cite{Iacobucci:2021ukp}.} \cite{Iacobucci:2021ukp}\cite{monpicoad}  was used to produce the PicoAD proof-of-concept prototype.  The ASIC  comprises a matrix of hexagonal pixels of 65 µm side (corresponding to  approximately 100 µm pitch), with an inter-pixel distance of 10 µm. Four of the pixels are connected to charge amplifiers with analog drivers, which are used to measure precisely the charge produced in the sensor. Figure \ref{fig:floorplan} shows the layout of the monolithic silicon pixel ASIC.

%% file: 3_TCAD_Simulations.tex
\section{TCAD simulations}
\label{sec:TCAD}

\begin{figure}[htbp]
\centering % \begin{center}/\end{center} takes some additional vertical space
\includegraphics[width=0.7\textwidth,trim=0 85 0 0,clip]{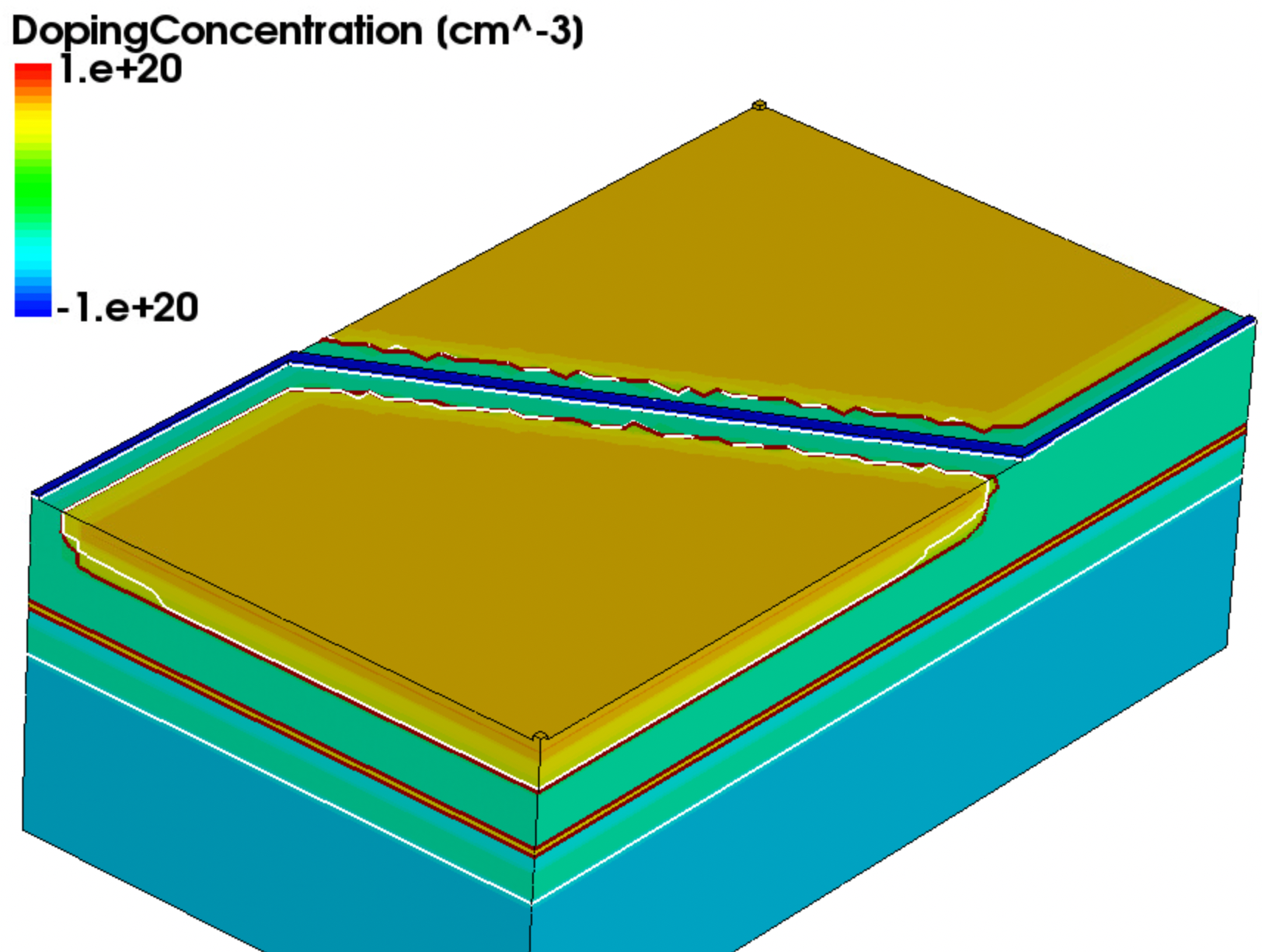}
% "\includegraphics" from the "graphicx" permits to crop (trim+clip)
% and rotate (angle) and image (and much more)
\caption{\label{fig:3Dgeo} TCAD 3D model geometry used for simulation of the PicoAD gain with fully depleted sensor. The volume simulated corresponds to one quarter of two adjacent hexagonal pixels.}
\end{figure}

A study of the expected behavior of the PicoAD proof-of-concept  prototypes was carried out using Technology Computer Aided Design (TCAD) simulations. The results refer to a 3D geometry model comprising a quarter of the hexagonal pixel with mirroring boundary conditions, which was used to assess the performance of the sensor at full depletion. Figure \ref{fig:3Dgeo} shows the model geometry.
Quasi-stationary ramp-up of the sensor bias was used to simulate the electric field in the sensor after full depletion of the epitaxial layer (comprising the gain layer). Such simulations show the stationary currents in the sensor, but they do not show the dynamic behavior associated to the removal of free carriers inside the sensor during ramp-up, which will be discussed in Section \ref{sec:probestation} using measurement data. 

\begin{figure}[htbp]
\centering % \begin{center}/\end{center} takes some additional vertical space
\includegraphics[width=.90\textwidth,trim=0 0 0 0,clip,angle=0]{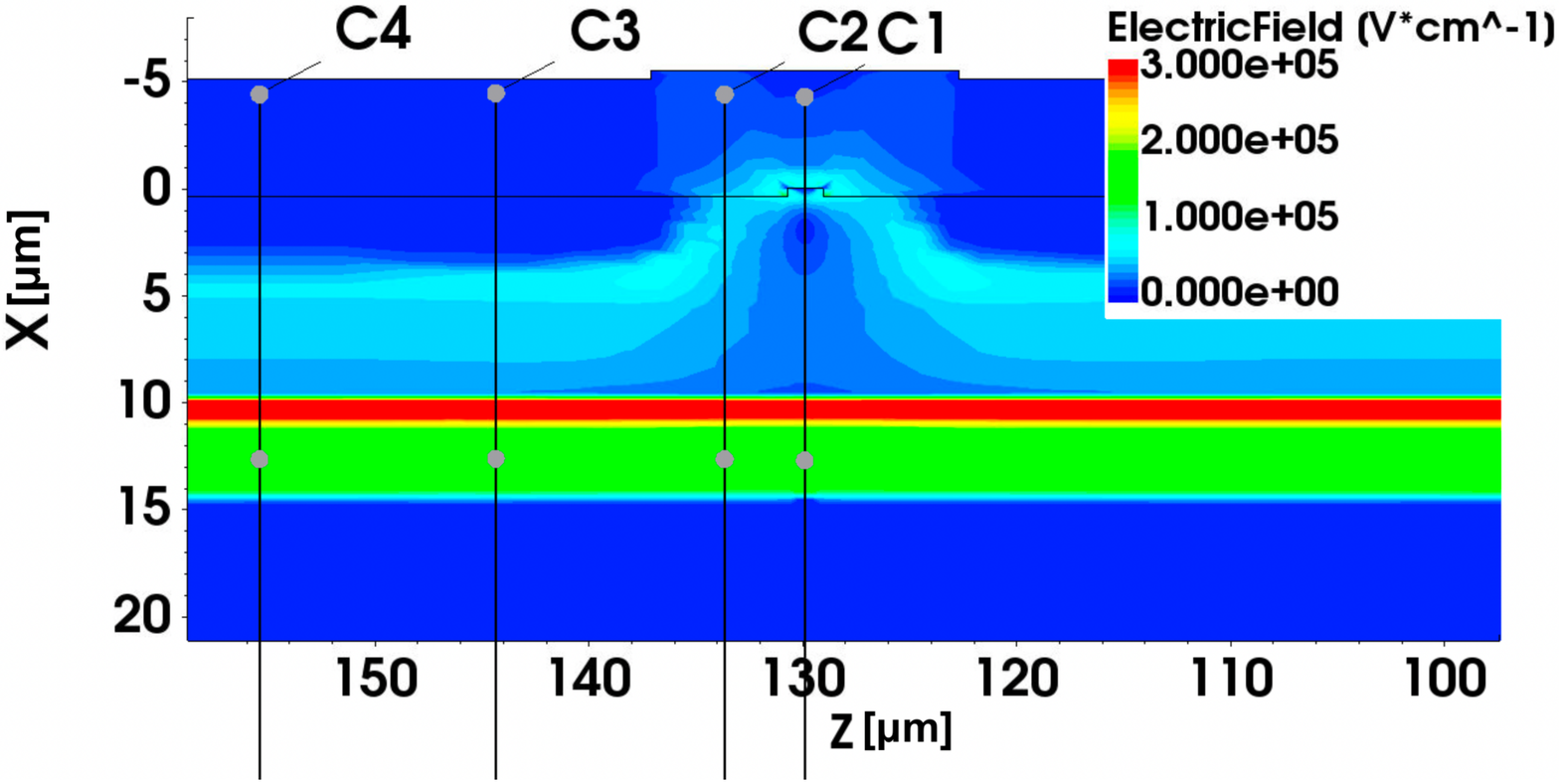}
\includegraphics[width=.80\textwidth,trim=0 0 0 0,clip,angle=0]{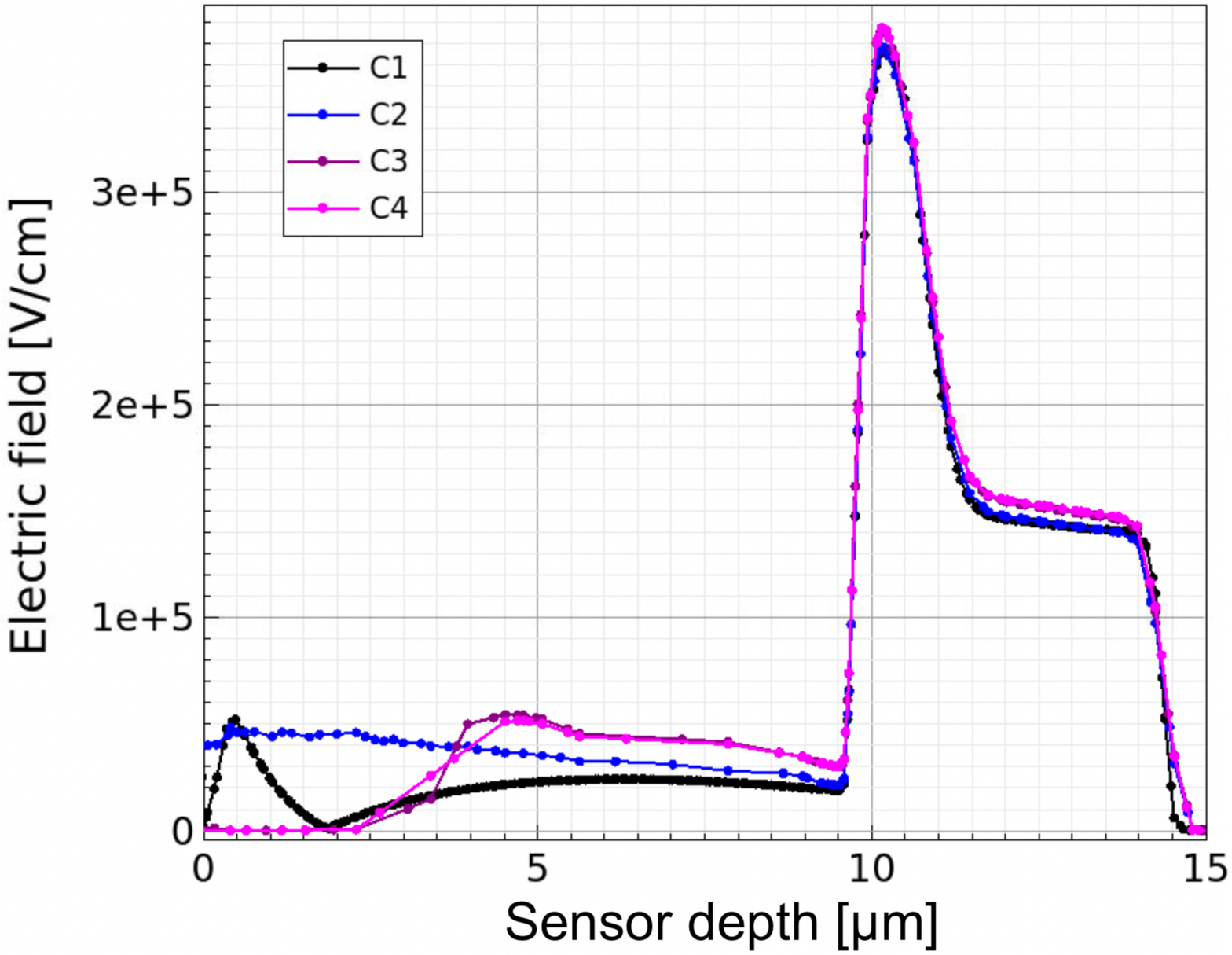}
% "\includegraphics" from the "graphicx" permits to crop (trim+clip)
% and rotate (angle) and image (and much more)
\caption{\label{fig:2DEfield} (Top) 2D cut of the modulus of the electric field in the sensor volume after full depletion at a bias voltage 2.5 V before breakdown. The vertical lines C1, C2, C3 and C4 represent the regions used for the 1D cuts used for the study of the electric field: C1 is at the center of the inter-pixel region; C2 at the edge of the collection electrode; C3 is close to the side of the collection electrode; C4 is at the center of the collection electrode. (Bottom) 1D cut of the modulus of the electric field in the four regions of the sensor volume. }
\end{figure}

Figure \ref{fig:2DEfield} shows the expected electric field intensity for a fully depleted PicoAD sensor at the highest implantation dose of the gain layer (dose 4).
At full depletion, the maximum value of the electric field is observed in the gain layer, with a small drop in the inter-pixel region due to the limited thickness of the drift region and to the presence of the P+ implant (PSTOP) used to stop the formation of a conductive channel under the shallow trench isolation. The field in the drift region, despite showing a gradient due to the relatively low resistivity of the epitaxial layer, is high enough to guarantee saturation of the electron drift velocity over a large fraction of the pixel cell while, as it is the case for this prototype, it is close to saturation even in the region under the PSTOP.

\begin{figure}[htbp]
\centering % \begin{center}/\end{center} takes some additional vertical space
\includegraphics[width=.60\textwidth,trim=0 0 0 35,clip]{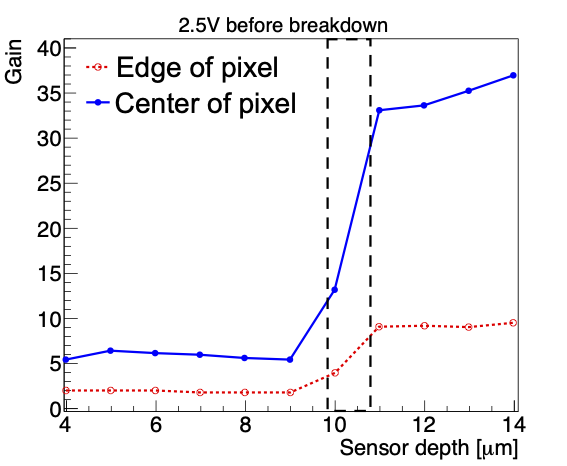}
% "\includegraphics" from the "graphicx" permits to crop (trim+clip)
% and rotate (angle) and image (and much more)
\caption{\label{fig:GainMap} Simulation of the gain produced by 63 electrons in a dose 4 PicoAD proof-of-concept prototype as a function of the initial position of the primary charge. The simulation is performed for 1 µm segments of a MIP in two of the cross sections reported in Figure \ref{fig:2DEfield}: C1, edge of the pixel (red dotted line); C3, center of the pixel (blue line). In this prototype, the gain layer is placed at 10 µm depth. The black dashed lines indicate the high field region produced by the gain layer. For a primary electron-hole position between 0 and 10 µm the avalanche is initiated by holes, between 10 and 14 µm the avalanche is initiated by electrons.}
\end{figure}

Transient simulations were performed to estimate the avalanche gain inside the sensor at a bias voltage 2.5 V below the gain layer breakdown. For this study, a primary charge of 63 electrons, corresponding to the most probable charge deposited by a MIP in one micron of silicon, was deposited on one-micron-long segments at different depths of the depletion region and the gain was evaluated as the ratio between the collected charge and the primary charge.  Figure \ref{fig:GainMap} shows the results of the simulation at the center and at the edge of the pixel (positions C3 and C1 of Figure \ref{fig:2DEfield}, respectively).

According to  design, the PicoAD gain layer amplifies the signal of electrons produced in the absorption region (electron gain), while it produces little to no gain when the primary charge is generated in the drift region (hole gain). It must be noted that  the small differences in electric field between the edge and the center of the pixel observed in this prototype  (see Figure \ref{fig:2DEfield}) produce a large difference in gain. The impact of these electric-field variations is more severe when the sensor is operated close to breakdown. Despite this, the simulation confirms that the presence of intense electric field in the gain layer both under the pixel and in the inter-pixel region shows that the sensor  provides 100\% gain fill-factor. 
Reduction of the ratio between the inter-pixel distance and the epitaxial layer thickness above the gain layer in future PicoAD prototypes will allow  better uniformity of the sensor gain.

%Figure \ref{fig:edgeeffect} shows the result of the quasi-stationary ramp-up of the bias voltage with the 2D model of the chip edge. The conductive behavior of the edge of the chip creates a steady current path between HV and the pixel if the gain layer implantation is in contact or too close to the edge itself as soon as the depletion layer reaches the gain implant. 
%Comparing this current at different regimes (Figure \ref{fig:edgeeffect}) shows that the current shifts from the pixel nwell to the innermost guard ring when the sensor bias increases. This condition suggests that the prototype sensor can be operated without incurring in loss of performance as long as the current is low enough to avoid damaging the silicon.

%\begin{figure}[htbp]
%\centering % \begin{center}/\end{center} takes some additional vertical space
%\includegraphics[width=.80\textwidth,trim=0 0 0 0,clip]{./Figures/IV_sim_edgeeffect_lowvoltage.png}
%\includegraphics[width=.80\textwidth,trim=0 0 0 0,clip]{./Figures/IV_sim_edgeeffect_highvoltage.png}
% "\includegraphics" from the "graphicx" permits to crop (trim+clip)
% and rotate (angle) and image (and much more)
%\caption{\label{fig:edgeeffect} \textcolor{red}{placeholder, need result from new model sim} Current injected from conductive edge due to insufficient separation between the gain layer and the chip edge. (top) During depletion the current is injected inside the pixel matrix. (bottom) in operation the extra current is collected by the guard ring, guaranteeing the correct behavior of the pixel matrix.}
%\end{figure}

%% file: 4_Probe_Station.tex
\section{Probe station measurements}
\label{sec:probestation}

\begin{figure}[htbp]
\centering % \begin{center}/\end{center} takes some additional vertical space
\includegraphics[width=.5\textwidth,trim=0 0 0 0,clip]{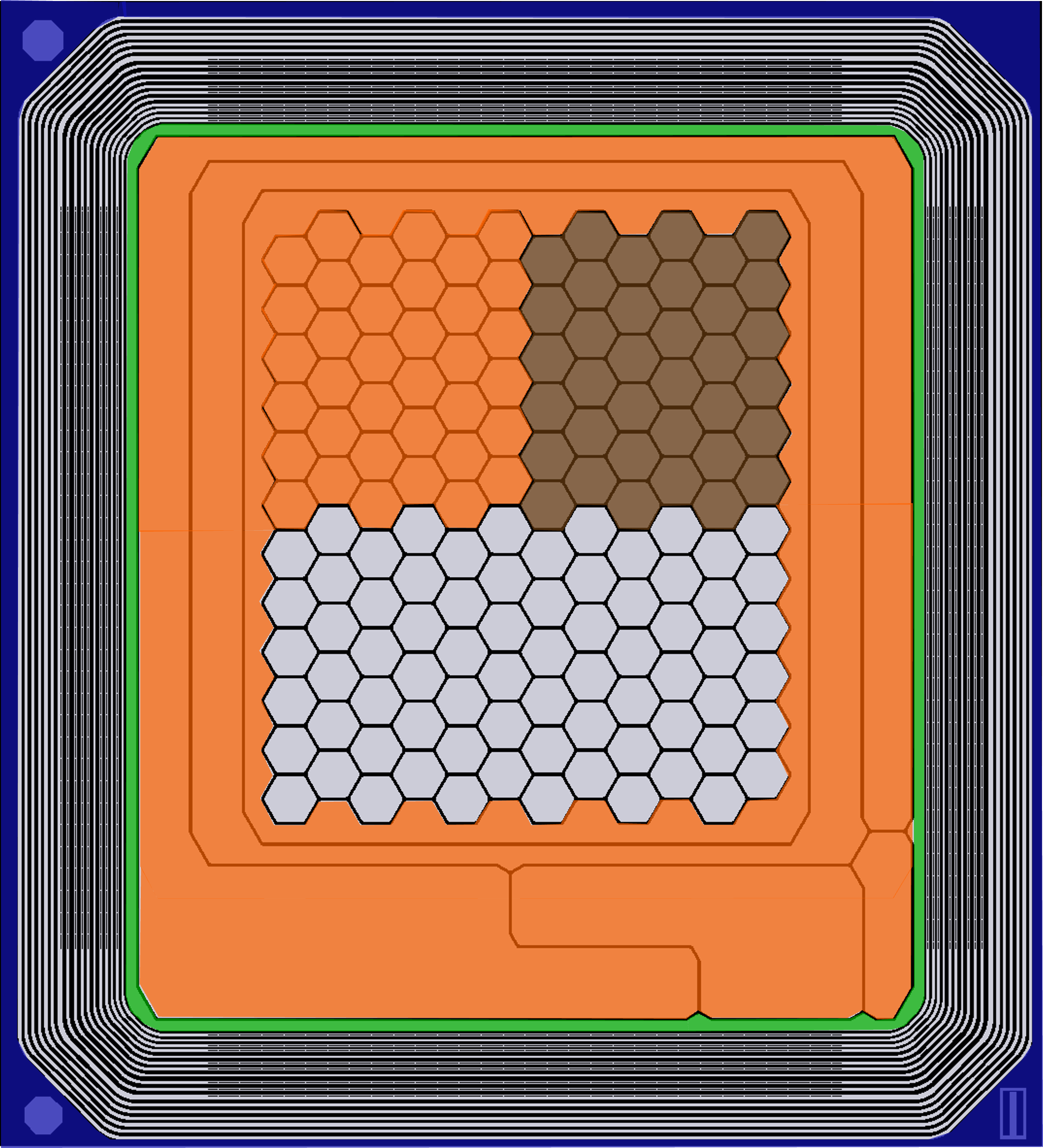}
% "\includegraphics" from the "graphicx" permits to crop (trim+clip)
% and rotate (angle) and image (and much more)
\caption{\label{fig:connections} The regions of the chip probed in the IV scans. Each color corresponds to a different region for which the current was measured independently. The brown region is one of the matrices of passive pixels. The orange region is the deep-Nwell that  hosts the electronics of the active pixels and  the periphery of the chip. The green region probes the current collected by the innermost guard ring. The blue region is the substrate, which was accessible from outside the guard ring on the top surface and from the backside connection. The pixels in grey and the guard rings were left floating.} 
\end{figure}

The IV characteristics of the four gain-layer doping doses of the PicoAD proof-of-concept prototypes were measured with the Cascade Microtech CM300 probe station of the cleanrooms of the University of Geneva. For the measurements, the substrate was connected to negative high-voltage, while the deep N-wells forming the pixel matrix and hosting the electronics were connected to ground. The probe station was able to measure the total current (\textit{substrate current}), but also  independently the current entering into a sub-matrix of pixels (\textit{pixel current}), the electronic N-wells (\textit{power current}) and the innermost guard-ring (\textit{Guard Ring current}).  Figure \ref{fig:connections} shows the location on the ASIC surface of the different N-wells as well as the  color-code associated to the current readout that will be used in below.

\begin{figure}[htbp]
\centering % \begin{center}/\end{center} takes some additional vertical space
\includegraphics[width=.49\textwidth,trim=4 4 4 4,clip]{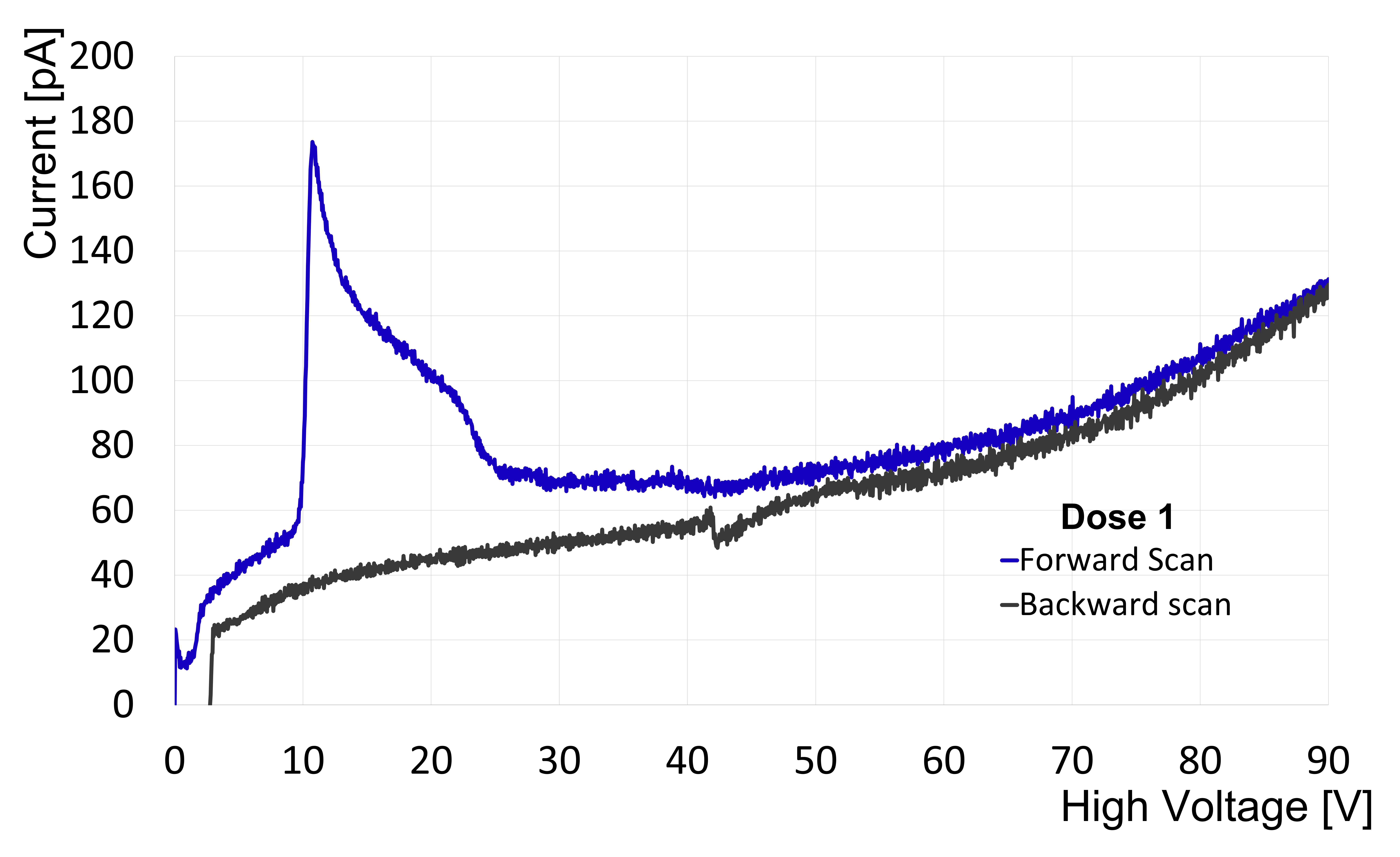}
\includegraphics[width=.49\textwidth,trim=4 4 4 4,clip]{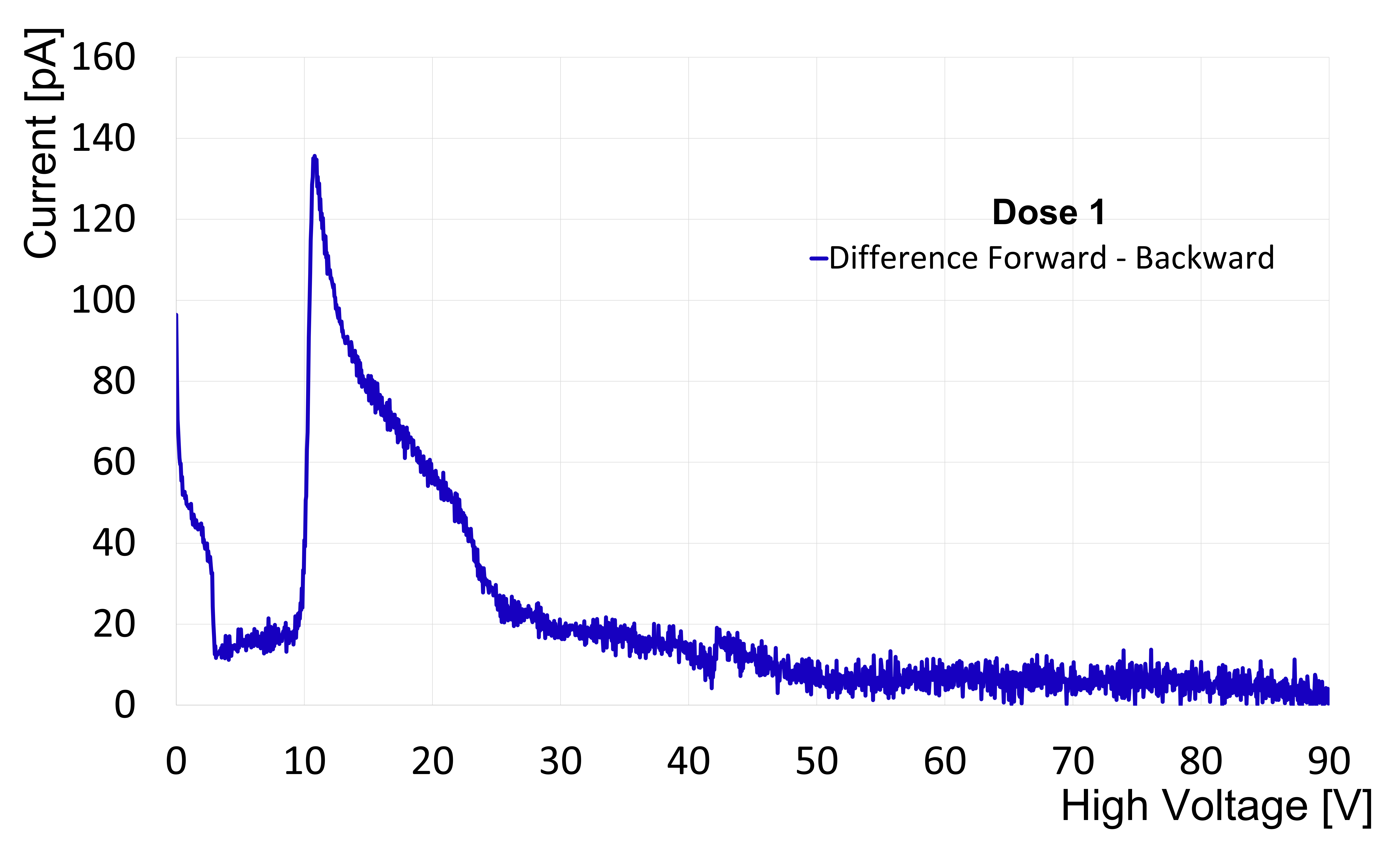}
% "\includegraphics" from the "graphicx" permits to crop (trim+clip)
% and rotate (angle) and image (and much more)
\caption{\label{fig:IV6} (Left) Substrate current as a function of the sensor bias voltage applied to the substrate of a PicoAD sensor with gain-layer implant dose 1. In blue the current measured  during the ramp up of the high voltage, in black the current during the ramp down. The reason for the current peak visible between 10 V and 30 V during ramp up, and not observed during ramp down of the high voltage, is given in the text. Each point is obtained averaging the current for 0.2 seconds. (Right) Current difference between ramp up and ramp down of the sensor bias voltage.} 
\end{figure}

%The prototypes were functional. 
Figure \ref{fig:IV6} left shows the substrate current as a function of the sensor bias  for a prototype chip with  dose 1 implant of the gain layer when a negative high voltage from 0 to 90 V was applied and subsequently ramped down. The sensor breakdown voltage was measured to be 185 V, and thus is not visible in Figure \ref{fig:IV6}. The substrate current deviates from a typical monotonic rise, showing a steep increase at 10 V during ramp-up, followed by a reduction and a return to the original trend line at 30 V. This behavior is not present when the voltage is ramped back down. The difference between ramp-up and ramp-down currents is displayed in Figure \ref{fig:IV6} right. The excess current between 10 and 30 V, which cannot be reproduced by TCAD quasi-stationary simulations, can be explained by the presence of free carriers inside the junction formed by the gain layer. Indeed, as soon as the high voltage is applied, the depletion layer starts extending downwards from the pixel junction into the epitaxial layer. During this phase the epitaxial layer above and below the N-type gain-layer implant is referred to negative high voltage, leaving the gain implant isolated with the free charges trapped inside the undepleted N-type gain layer. The removal of these trapped charges produces the measured non-stationary current, that disappears when the PicoAD is fully depleted. The magnitude of the extra current depends on the speed of the high voltage ramp, since it is produced by the removal of a fixed amount of charge.

\begin{figure}[htbp]
\centering % \begin{center}/\end{center} takes some additional vertical space
\includegraphics[width=.49\textwidth,trim=4 4 4 4,clip]{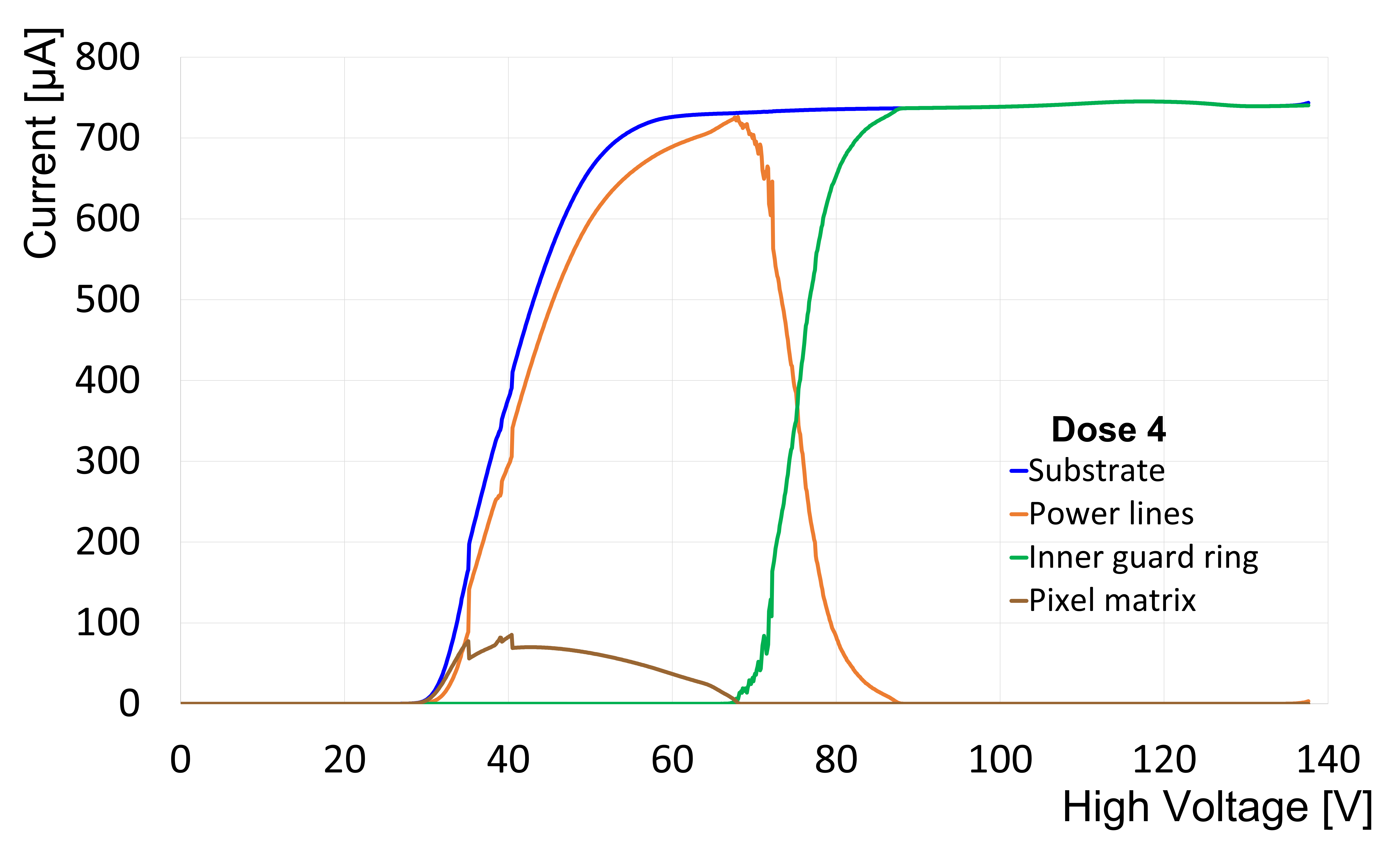}
\includegraphics[width=.49\textwidth,trim=4 4 4 4,clip]{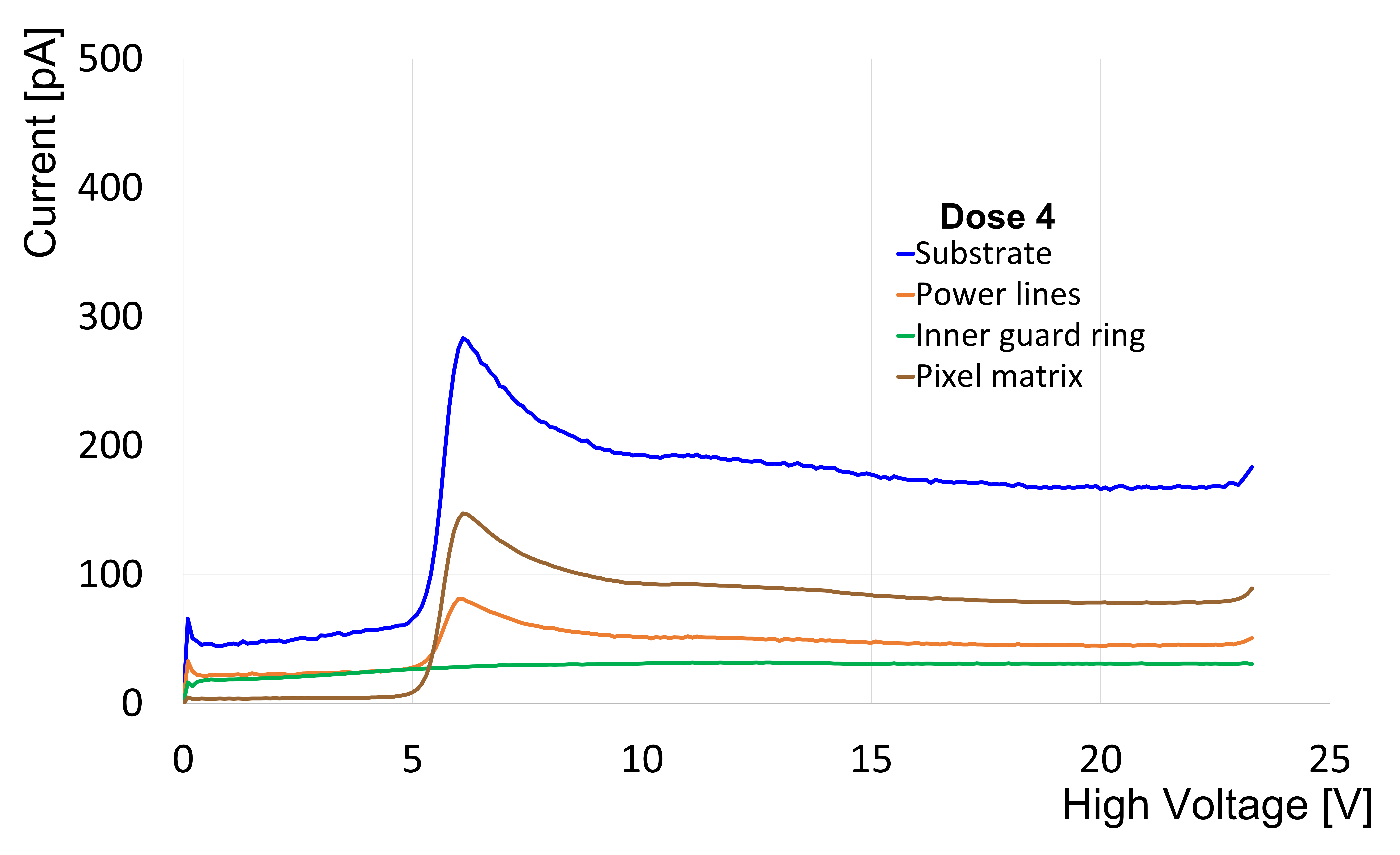}
% "\includegraphics" from the "graphicx" permits to crop (trim+clip)
% and rotate (angle) and image (and much more)
\caption{\label{fig:IV9} IV measurement of a PicoAD sensor with gain-layer implant dose 4. (Left) In blue: the total current collected by the substrate. In brown: the current collected by the pixel terminals. In orange: the current collected by the power lines. In green: the current collected by the innermost guard ring. When the full depletion of the sensor is reached all the extra current is collected by the guard ring. (Right) Current collected by the same terminals before the onset of the extra current observed beyond 25 V. The effect of the depletion of the gain layer under the pixels and power lines is visible above 5 V.
} 
\end{figure}

Figure~\ref{fig:IV9} shows the IV measurement performed on a sample with the highest gain-layer implant dose (dose 4) from the same wafer. In this case, a large extra current was visible after reaching the start of the depletion of the gain layer. The origin of this extra current is attributed to the region outside the guard ring (possibly the chip edge). TCAD simulations suggest that this current could be injected to the pixels and power lines via the undepleted gain layer, which is the case of PicoAD sensors with the gain layer extending under the guard ring. This extra current was mostly observed for samples with larger gain-layer implant dose and with an early onset of the floating n-layer depletion. The presence of local defects in silicon discussed in Section \ref{sec:chip} could contribute to the activation of this extra current. Figure \ref{fig:IV9} also shows that, after full depletion of the gain layer, all the excess current is collected by the guard ring and therefore this current has no influence on the sensor performance. This effect can be avoided by removing the part of the gain-layer implant that is underneath the guard ring.

%% file: 5_Climate_Chamber.tex
\section{Measurements with a $^{55}$Fe radioactive source}
\label{sec:iron}

The response in gain of the proof-of-concept PicoAD prototype sensors was studied using a $^{55}$Fe source.  The measurements were performed in a climate chamber at an ambient temperature between  -20 and +20 $^{\circ}$C. The output of the analog pixels were sent to an oscilloscope with 1 GHz analog bandwidth and the waveforms were acquired for off-line analysis.

\begin{figure}[htbp]
\centering % \begin{center}/\end{center} takes some additional vertical space
\includegraphics[width=.52\textwidth,trim=0 0 0 0,clip]{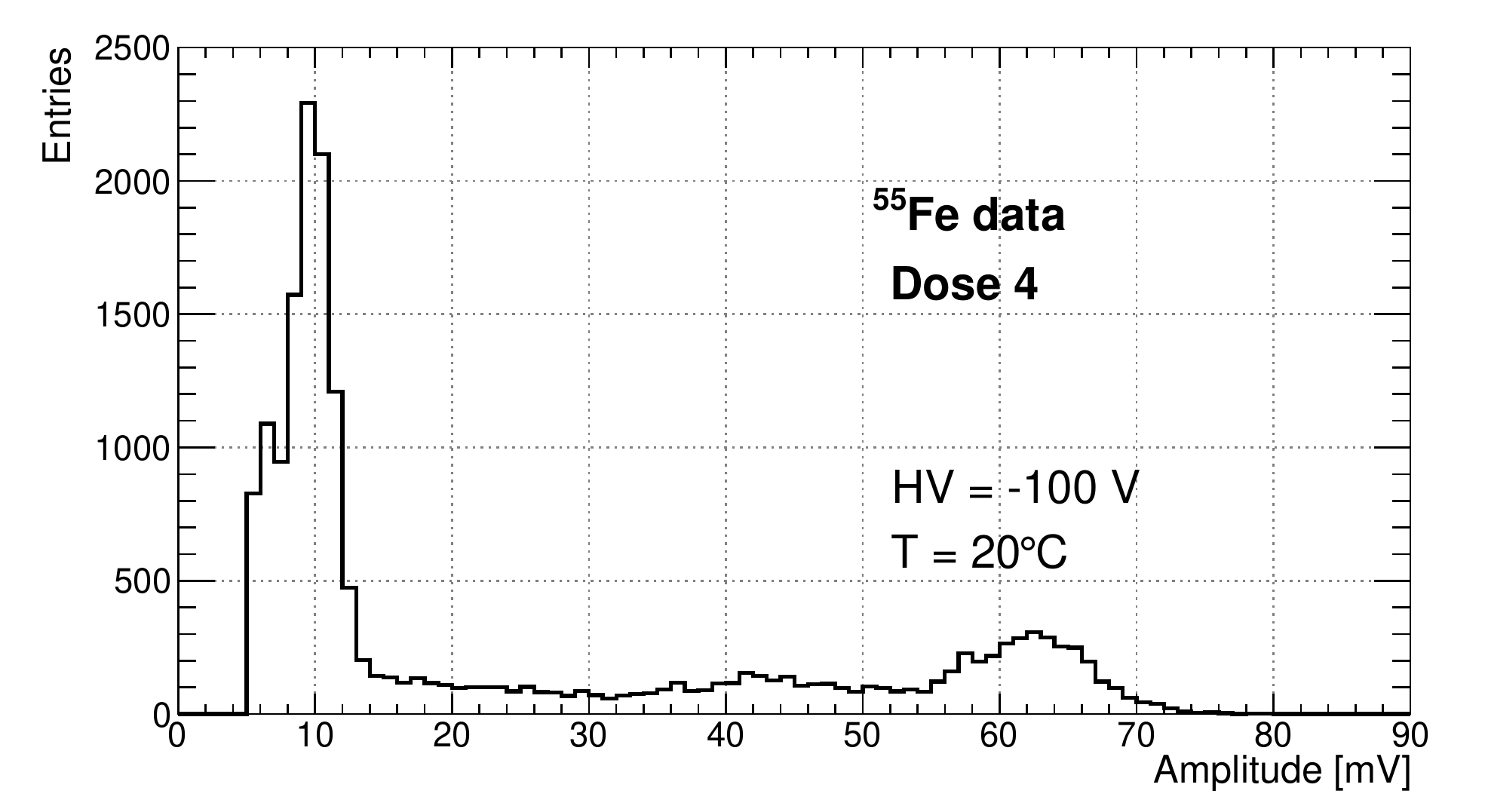}
\includegraphics[width=.47\textwidth,trim=10 0 0 0,clip]{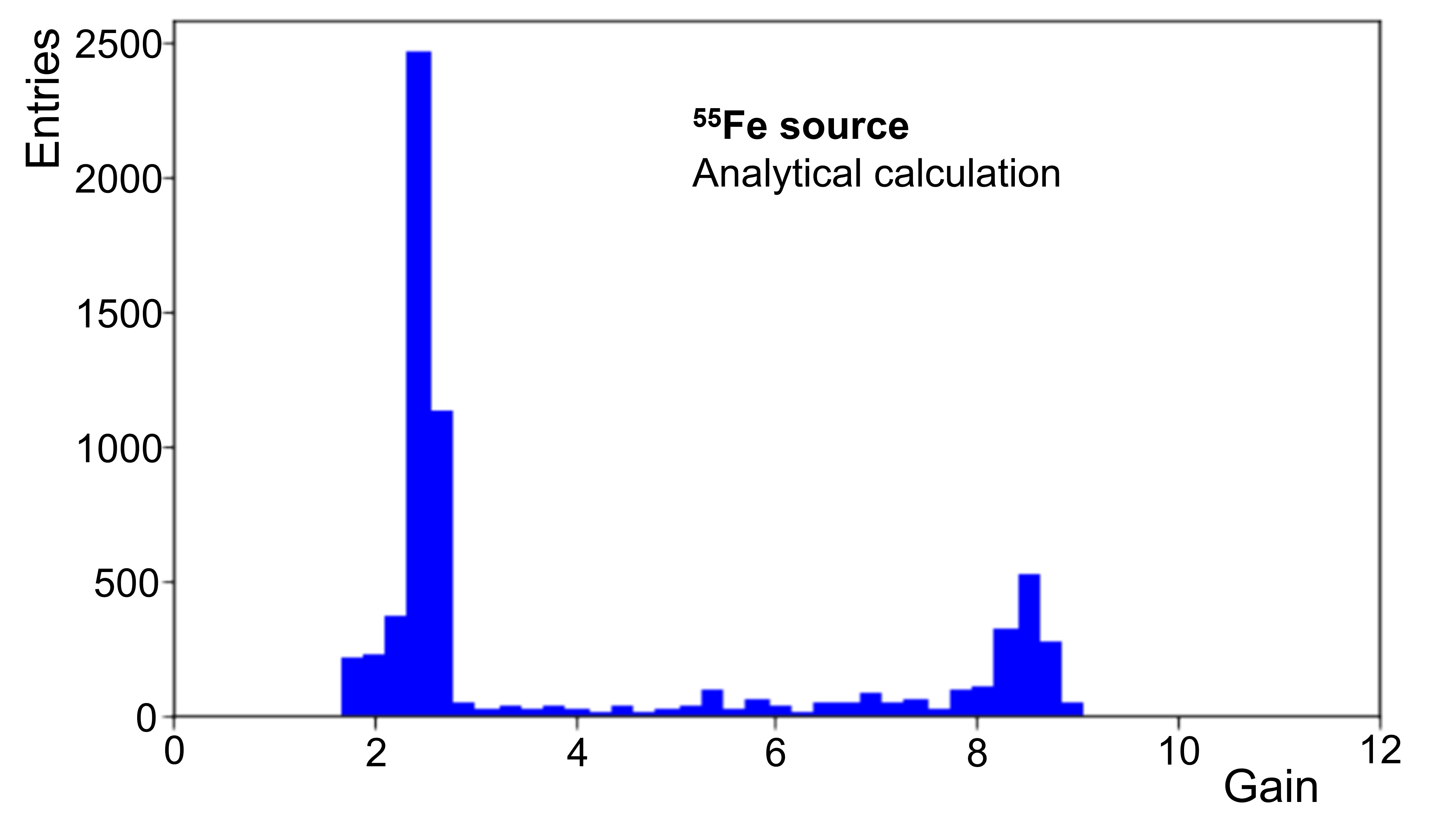}
% "\includegraphics" from the "graphicx" permits to crop (trim+clip)
% and rotate (angle) and image (and much more)
\caption{\label{fig:spectrum} (Left) Signal amplitude spectrum measured with a $^{55}$Fe source for a dose4 PicoAD sample at a temperature of +20 $^{\circ}$C and high voltage of 100 V. ~(Right) Gain spectrum obtained using the electric field profiles from TCAD simulation and the analytical formula given in~\cite{MAES1990705} for the gain obtained from impact ionization.}
\end{figure}

The 5.9 keV photons emitted by the $^{55}$Fe X-ray source interact via photoelectric effect in the silicon, producing a charge cluster inside the sensor that can be approximated as a point-like charge deposition of 1640 electrons.
Figure \ref{fig:spectrum} left shows an example of the signal amplitude distributions obtained with the $^{55}$Fe source from a dose 4 sample. 
The peak in amplitude visible at 10 mV can be attributed to X-ray photons that convert inside the PicoAD drift region; in this case, holes will drift towards the gain region and 
produce {\it hole gain}.
Conversely, the peak at 45 mV can be attributed to photon conversions within the thin absorption region; in this case, the primary electrons will drift towards the gain layer, producing the much larger {\it electron gain}.
%The two peaks at 10 and 45 mV  are produced by photons that converted inside the PicoAD drift and absorption regions, respectively: when the conversion happens in the drift region, holes are injected into the gain layer, producing hole gain; when the X-ray photon converts in the absorption region electrons are injected into the gain layer, producing electron gain. 
The flat region of the spectrum between the two peaks can be associated  to  photons that convert either in the inter-pixel area, where the gain is lower, or inside the gain layer. 
For a qualitative comparison, the same spectrum was produced  using the electric field profile extracted from TCAD simulations and calculating the gain with the impact multiplication formula of~\cite{MAES1990705} for primary clusters uniformly distributed in the sensor volume. The result, shown in Figure~\ref{fig:spectrum} right, supports the interpretation given above.

The gain calculation shown in Figure~\ref{fig:GainMap} obtained with TCAD transient simulation also corroborates this interpretation. 

\begin{figure}[htbp]
\centering % \begin{center}/\end{center} takes some additional vertical space
\includegraphics[width=.49\textwidth,trim=0 0 0 0,clip]{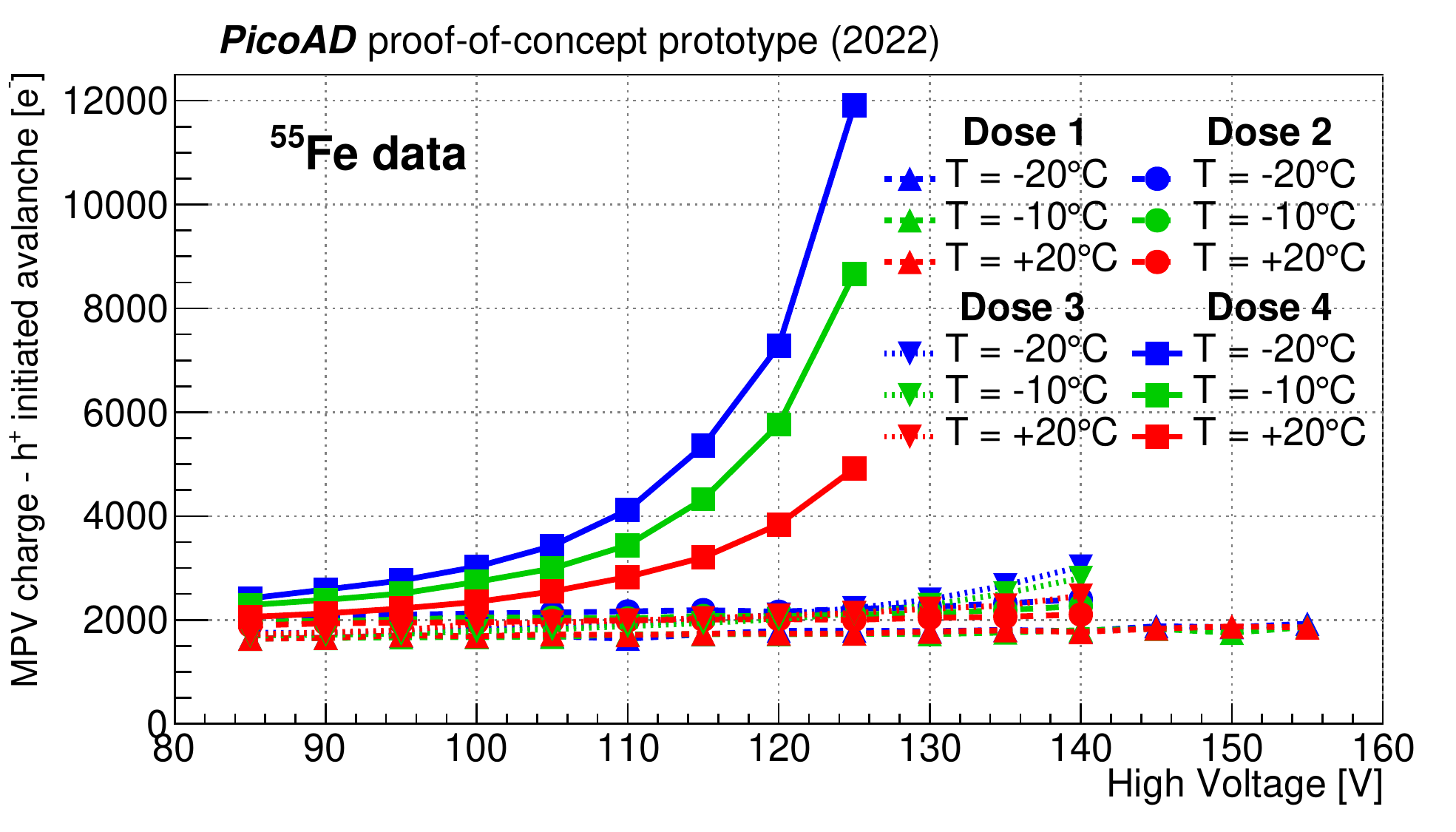}
\includegraphics[width=.49\textwidth,trim=0 0 0 0,clip]{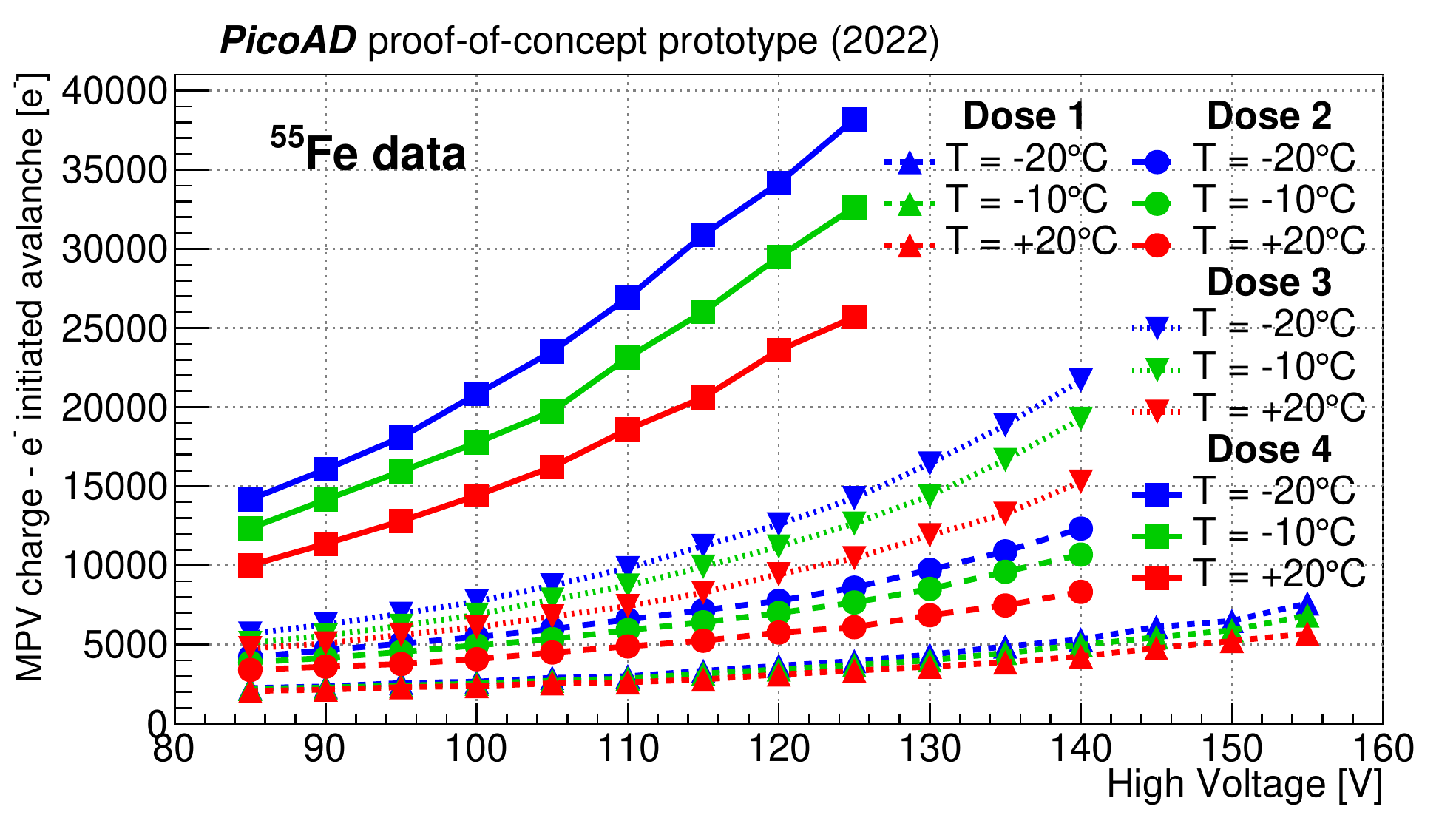}
% "\includegraphics" from the "graphicx" permits to crop (trim+clip)
% and rotate (angle) and image (and much more)
\caption{\label{fig:peaks} Average collected charge of the signals produced in the PicoAD by the conversion of X-ray photons from a $ ^{55}$Fe source. (Left) with hole gain produced by photon conversion in the drift region. (Right) with electron gain produced by photon conversions in the absorption region.
}
\end{figure}

The mean amplitudes associated to the two peaks were extracted with Gaussian fits. The values were normalized by the gain of the charge amplifier to obtain the collected charge. Figure \ref{fig:peaks} shows the average collected charge for the hole-gain (left) and electron-gain (right) peaks of the $^{55}$Fe spectrum as a function of the HV for the four gain-layer implant doses, measured at temperatures between -20 and +20 $^{\circ}$C. All sensors show an increase of the signal mean collected charge with increasing sensor bias, which indicates that electron (hole) gain is present for charge clusters produced in the absorption (drift) region. 

\begin{figure}[htbp]
\centering % \begin{center}/\end{center} takes some additional vertical space
\includegraphics[width=.75\textwidth,trim=0 0 0 0,clip]{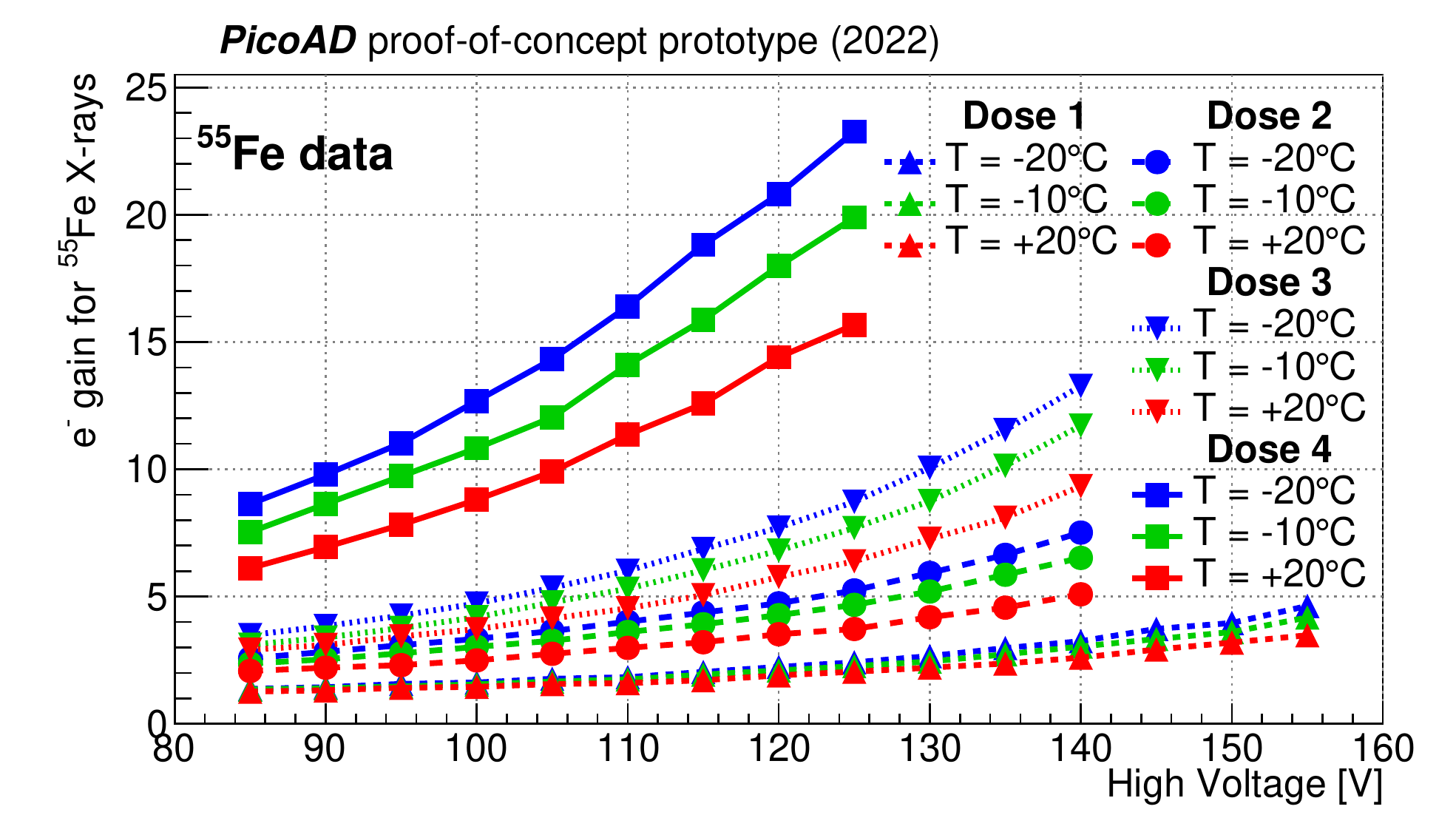}
% "\includegraphics" from the "graphicx" permits to crop (trim+clip)
% and rotate (angle) and image (and much more)
\caption{\label{fig:gain} Gain measured for the conversion of X-ray photons from a $^{55}$Fe source for electron-initiated avalanche as a function of the sensor bias voltage. For each temperature, the gain is obtained as the ratio between the mean collected charge of signals generated by 5.9 keV photon conversion in the PicoAD absorption layer and the mean value obtained at 85 V for conversions in the drift region in the sensor with gain-layer dose 1.
}
\end{figure}

To extract the sensor gain, it was assumed that for each measured temperature the average charge of the first (hole) peak at the lowest measured bias voltage of 85 V and for the lowest gain layer doping concentration (dose 1), corresponds to the signal charge expected in absence of avalanche gain. This value was then used to normalize the average charges of the second (electron) peak at different bias voltages to estimate the gain for avalanches initiated by electrons in the absorption region. Figure \ref{fig:gain} shows the result. 

A maximum electron gain of 23 was observed for the sensor with gain-layer implant dose 4, operated at the lowest ambient temperature of -20 $^{\circ}$C. Coherently with the expectation,  sensors with  lower gain-layer implant dose show lower electron gain at a fixed bias voltage.

\begin{figure}[htbp]
\centering % \begin{center}/\end{center} takes some additional vertical space
\includegraphics[width=.80\textwidth,trim=0 0 0 0,clip]{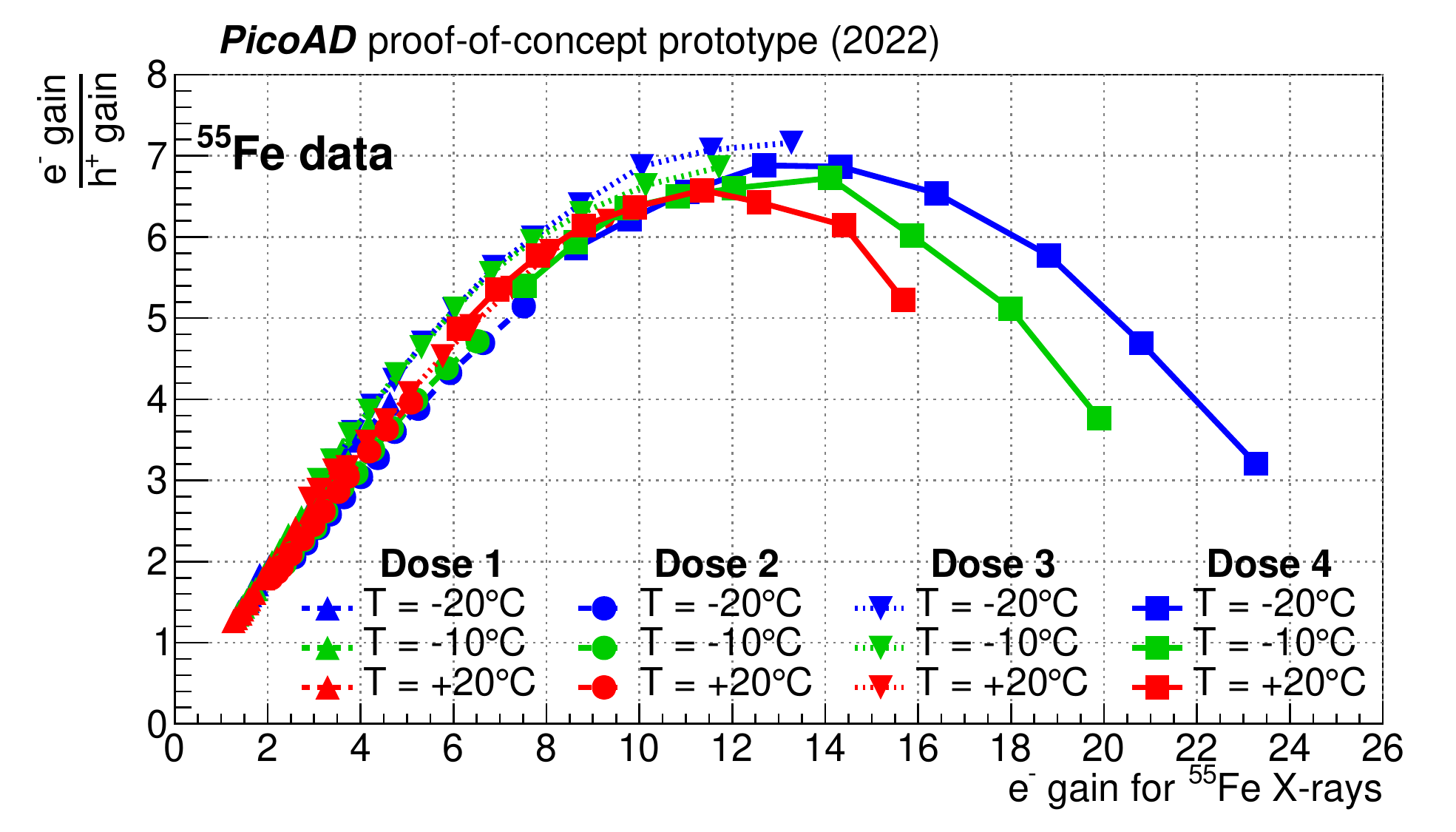}
% "\includegraphics" from the "graphicx" permits to crop (trim+clip)
% and rotate (angle) and image (and much more)
\caption{\label{fig:gainratio} Ratio between the electron gain and the hole gain obtained with a $^{55}$Fe source for the four devices tested, as a function of the electron gain.}
\end{figure}

%To further study the characteristic of the gain layer, 
The PicoAD structure offers the opportunity to study the relation between the electron and hole gain in the same junction, which can be used to infer the properties of the avalanche. Figure \ref{fig:gainratio} shows the ratio between electron gain and hole gain as a function of the electron gain. The hole gain was computed with the same method used for the electron gain and described above. This representation of the data removes the direct dependence of the gain on the temperature and the sensor bias, showing only the characteristics of the impact multiplication in silicon, that in turn depends on parameters of the gain layer, such as its width or the magnitude of the electric field. 

As expected, all the devices are on the same trend curve, since the implantation energy used for the different doses are the same. For the sensor with dose 4, which shows the highest absolute electron gain, a reduction of the electron/hole gain ratio is measured for electron gain larger than 12. This observation cannot be reproduced by the model of the impact ionization with any parametrization for the impact parameters \cite{MAES1990705}. 

The reduction of the gain ratio at large values of electron gain can be explained by transient space-charge effects during the development of the avalanche, which would reduce the electric field in the gain layer at high gain when the primary cluster is produced in the absorption region. The time-dependent spatial charge density varies with  the total charge drifting inside the sensor, which is the product of the gain and the primary charge. The same TCAD simulation introduced in Section \ref{sec:TCAD} to obtain the gain profile inside the sensor can be used to estimate the impact of  space-charge effects on the measured electron gain. 
\begin{figure}[htbp]
\centering % \begin{center}/\end{center} takes some additional vertical space
\includegraphics[width=.60\textwidth,trim=0 0 0 42,clip]{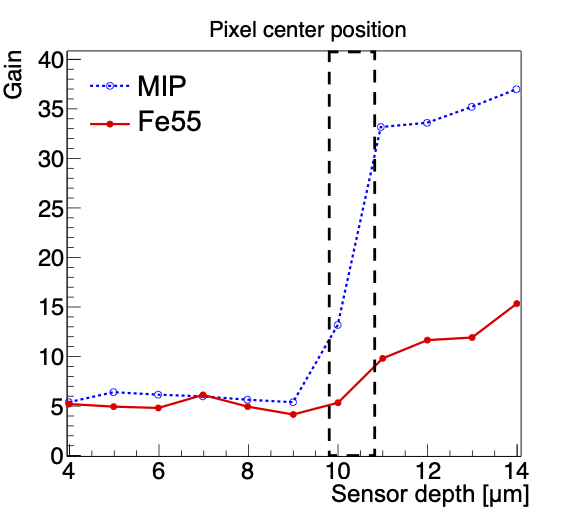}
\caption{\label{fig:spacecharge} TCAD simulation of the gain  of a point-like primary charge deposition at the center of a PicoAD pixel and at varying depth of 63 electrons (MIP, blue dotted curve, also shown in Figure~\ref{fig:GainMap}) and 1640 electrons ($ ^{55}$Fe X-ray conversion, red curve). The black dashed lines indicate the high field region produced by the gain layer.}
\end{figure}
Figure \ref{fig:spacecharge} compares the gain profile expected for a primary charge deposited at different sensor depths at the center of the pixel. Primary charges of 63  and 1640 electrons were used, the first representing the ionisation of a MIP in 1 µm and the latter  the charge deposited by photoelectric effect by an X-ray from the $ ^{55}$Fe source.
The simulation confirms that the  large electric-charge density present in the case of $^{55}$Fe  X-ray photon conversion  limits considerably the gain for an avalanche initiated by electrons, while it has little to no impact when the avalanche is initiated by holes. This result might explain the behavior observed at high electron gain in Figure \ref{fig:gainratio}. It also suggests that the actual electron gain achievable with MIPs is much larger than the value measured with the $^{55}$Fe source.

%% file: 6_Conclusions.tex
\section{Conclusions}
\label{sec:discussion}

Proof-of-concept prototypes of the monolithic PicoAD were produced at IHP using 5 µm (absorption) + 10 µm (drift) thick epitaxial layers. The prototypes are functional. Probe station transient IV measurements show a depletion of the gain layer compatible with expectations.

Measurements with a $ ^{55} $Fe X-ray source show a dependence of the electron gain with the dose of the gain layer compatible with the expectations. The highest electron gain measured with $ ^{55} $Fe source is 23. The PicoAD structure allows measuring electron and hole gain on the same device. The ratio between the electron gain and hole gain deviates from the trend predicted by the model for the charge multiplication produced by impact ionization. This deviation is compatible with a reduction of the electric field during the development of the electron-initiated avalanche for a high-density primary charge deposition. A TCAD transient simulation comparing the gain for a point-like charge deposition from a $ ^{55} $Fe X-ray source to the one expected for the charge density of a MIP corroborates this hypothesis.